\documentclass[11pt,a4paper]{ctexart}

\usepackage[a4paper,margin=1in]{geometry}
\usepackage{amsmath,amssymb,bm}
\usepackage{graphicx}
\usepackage{booktabs}
\usepackage{multirow}
\usepackage{array}
\usepackage{tabularx}
\usepackage{longtable}
\usepackage{caption}
\usepackage{subcaption}
\usepackage{authblk}
\usepackage{url}
\usepackage{xcolor}
\usepackage{cite}
\usepackage{hyperref}

\hypersetup{
	colorlinks=true,
	linkcolor=blue,
	citecolor=blue,
	urlcolor=blue
}

\newcommand{\safeincludegraphics}[2][]{%
	\IfFileExists{#2}{%
		\includegraphics[#1]{#2}%
	}{%
		\fbox{\parbox[c][0.28\textheight][c]{0.92\linewidth}{\centering Missing figure file: \texttt{#2}}}%
	}%
}

\title{Prediction of Celestial Pole Offsets Based on Sliding Window and Bivariate Least Squares Fitting}
\author[1,2]{Wang Wei-long}
\author[1,2]{Wu Yuan-wei}
\author[1]{Li Xi-shun}
\author[1,2]{Qiao Hai-hua}
\author[4]{Kong Qiao-li}
\author[1,2]{Yang Hai-yan}
\author[1,2,3]{Yang Xu-hai}
\affil[1]{National Time Service Center, Chinese Academy of Sciences, Xi'an 710600, China}
\affil[2]{University of Chinese Academy of Sciences, Beijing 100049, China}
\affil[3]{Key Laboratory of Time Reference and Applications, Chinese Academy of Sciences, Xi'an 710600, China}
\affil[4]{College of Geodesy and Geomatics, Shandong University of Science and Technology, Qingdao 266590, China}
\date{}

\begin{document}
	\maketitle
	
	\begin{abstract}
		As an important component of Earth Orientation Parameters (EOP), the prediction of Celestial Pole Offsets (CPO) holds significant importance for missions such as deep space exploration. To explore a better CPO prediction algorithm that improves accuracy while achieving generalization across different forecast spans, a CPO prediction algorithm is proposed based on a sliding window and bivariate least squares fitting. First, experiments determine an optimal sliding window of 900 days. Then, bivariate least squares fitting is performed on the selected 900-day historical data to complete extrapolation prediction. Experimental results show that the proposed algorithm exhibits excellent accuracy. In comparisons with prediction results from participating teams in the Second Earth Orientation Parameters Prediction Comparison Campaign (2nd EOP PCC), the algorithm's Mean Absolute Error (MAE) is superior to both ID154 and ID155. Team ID154 achieved the best dX prediction, while Team ID155 achieved the best dY prediction. Furthermore, the algorithm performs well not only on the EOP 14 C04 series but also on the newly released EOP 20 C04 series after the 2nd EOP PCC. Its prediction results are far better than those in the daily files published by the International Earth Rotation and Reference Systems Service (IERS). In terms of dX forecast accuracy, the MAE for the 10th, 30th, and 57th days were reduced by 53\%, 59\%, and 60\%, respectively. In terms of dY forecast accuracy, the MAE for the 10th, 30th, and 57th days were reduced by 35\%, 38\%, and 42\%, respectively.
	\end{abstract}
	
	
	\section{Introduction}
	
	Earth orientation parameters (EOP) are core physical quantities that describe the state of Earth's rotation and spatial attitude. They consist of day length, polar motion (PMX, PMY), and precession and nutation. As important parameters for the mutual conversion between the Terrestrial Reference Frame and the Celestial Reference Frame, EOP plays a crucial role in fields such as deep space exploration, satellite orbit determination, and satellite navigation. Currently, the measurement of EOP primarily relies on space geodetic technologies such as Very Long Baseline Interference (VLBI), Global Navigation Satellite System (GNSS), and Satellite Laser Ranging (SLR)~\cite{BehrendBaver2007IVSAnnualReport,RatcliffGross2010SPACE2008}. Since observed actual data require complex computation to obtain the actual EOP values, and real-time EOP values are needed for practical engineering tasks such as satellite navigation and deep space exploration, research on EOP prediction methods and prediction accuracy is of utmost importance.
	
	The definition of celestial pole offset is the deviation between the observed celestial pole (determined by VLBI technology) and the Celestial Intermediate Pole (CIP) predicted by the official model of the International Astronomical Union (IAU), such as IAU 2000A/2006. It is usually represented by two components, dX and dY, in a rectangular coordinate system. Currently, only VLBI technology can achieve high-precision measurement of this parameter~\cite{KianiShahvandi2024DeepEnsembleCPO}. However, VLBI data processing takes several weeks, resulting in a significant lag in the final CPO product, which can be up to 4 weeks~\cite{Belda2018ImproveCPOPrediction}. This lag is difficult to meet the real-time Earth orientation parameter requirements for applications such as spacecraft navigation and deep space telescope pointing. Therefore, conducting reliable CPO prediction research has become the key to solving this contradiction~\cite{KianiShahvandi2024DeepEnsembleCPO}.
	
	The VLBI data processing results after 2000 indicate that the time series of CPO contains rich information about the internal dynamics and external perturbations of the Earth. The main fluctuation component originates from Free Core Nutation (FCN)~\cite{MathewsHerringBuffett2002NutationPrecession}. FCN is a free rotation mode of the Earth, caused by the rotation of the liquid elliptical core within the viscoelastic mantle, with its rotation axis slightly deviating from the mantle's rotation axis. Observationally, in a spatially fixed reference frame, FCN manifests as a retrograde motion of the Earth's shape axis, opposite to the direction of Earth's rotation, with a period of about 430 days~\cite{LambertDehant2007CoreParametersVLBI}.
	
	In recent years, with the continuous deepening of research on CPO, many scholars have summarized various forecasting methods. Zhu Qiang et al. proposed a combined model of least squares and autoregression (LS+AR) for the short-term and medium-to-long-term forecasting needs of CPO sequences: first, the linear trend and main periodic term (FCN, −430.21 d) are fitted and extrapolated using least squares, and then an AR model is established for the residuals for extrapolation, resulting in the final prediction through superposition. Based on 200 rolling experiments on the IERS 08C04 sequence, this method significantly outperforms the internationally mainstream Lambert model in short-term forecasting accuracy (1--5 days); in the 5--90 day range, the RMS errors in the dX and dY directions stabilize at approximately 110~$\mu$as and 130~$\mu$as, respectively, representing an overall reduction of 20--40\% compared to the Lambert model ($\approx$120~$\mu$as, $\approx$220~$\mu$as). Especially, the improvement in the medium-to-long-term segment of the dY component is evident, providing a new approach for high-precision forecasting of Earth orientation parameters~\cite{ZhuXuZhou2015CPO_LS_AR}; Nastula et al. proposed a prediction method based on Kalman filtering and smoother, fitting the annual and FCN dual cycles within a 365-day sliding window, constructing a state space model by integrating historical observation data, and validating the predictions of the dX and dY components of the EOP C04 sequence. The results show that the average absolute error of the dX component in short-term forecasts (1--30 days) is reduced by 12\%--18\% compared to the traditional LS+AR model, effectively smoothing out observational noise disturbances~\cite{Nastula2020KalmanCPO}; Belda et al. proposed a method based on the FCN empirical model B16, extracting FCN parameters with a 400-day sliding window and extrapolating CPO. Data testing from 2000 to 2016 showed that the 40-day prediction error is approximately 85 µas, representing a 35\%--40\% improvement over the accuracy of IERS rapid service products~\cite{Belda2018ImproveCPOPrediction}; Kiani Shahvandi et al. further proposed a deep integrated geophysical information neural network model embedded with effective angular momentum (EAM) physical constraints. In data experiments from 2000 to 2020, the root mean square error for a 7-day prediction period is superior to that of the Bulletin A product, with a 23\% improvement in the accuracy of the dY component~\cite{KianiShahvandi2024DeepEnsembleCPO}.
	
	From September 2021 to December 2022, our team from the National Time Service Center participated in the 2nd EOPPCC event. We utilized the LS+AR method to forecast CPO. The forecast accuracy in the dX direction ranged from 61-93 uas within 0-30 days and from 84-116 uas within 31-88 days. In the dY direction, the forecast accuracy was between 91-127 uas for 0-30 days and between 115-148 uas for 31-88 days. During the 2nd EOPPCC event, the group with the best dX forecast results was ID154, achieving a forecast accuracy of 41-58 uas within 0-31 days. For the dY forecast, the best group was ID155, with an accuracy of 49-86 uas within 0-31 days. We identified several issues with the CPO forecasts:
	\begin{enumerate}
		\item Most teams find it difficult to achieve good forecasts for both dX and dY parameters;
		\item Due to the time-varying intensity of the FCN signal, this factor needs to be considered in the forecast to achieve better fitting of the data.
	\end{enumerate}
	To this end, based on the research methods of Belda et al., we propose a CPO forecasting method utilizing sliding windows and bivariate fitting.
	
	This article primarily consists of three parts. The first part is the introduction, which outlines the research background and significance of this article. The second part introduces the data and methods, specifying the data sources and describing how the data was processed. The third part presents our evaluation results and discussion, evaluating and validating the algorithm presented in this article, and drawing the final conclusion.
	
	\section{Data and Methods}
	
	CPO forecasting requires the use of EOP C04 provided by the International Earth Rotation and Reference Systems Service (IERS), as well as the daily products of rapid products and forecasting products. For details, please refer to \url{https://www.iers.org/IERS/EN/DataProducts/EarthOrientationData/eop.html}. It should be noted here that EOP files are updated daily, but CPO data is not updated daily, but rather 1-2 times per week.
	
\begin{figure}[htbp]
		\centering
		\begin{subfigure}[t]{0.48\linewidth}
			\centering
			\safeincludegraphics[width=\linewidth]{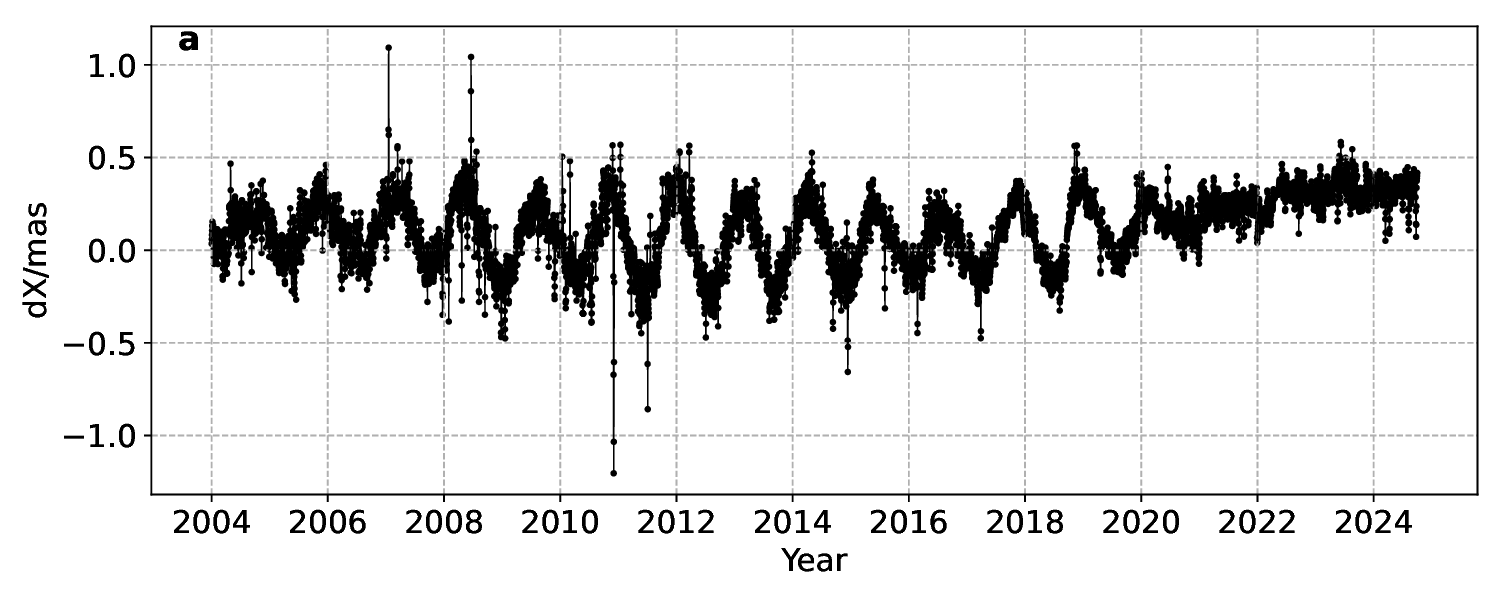}
			\caption{dX}
		\end{subfigure}\hfill
		\begin{subfigure}[t]{0.48\linewidth}
			\centering
			\safeincludegraphics[width=\linewidth]{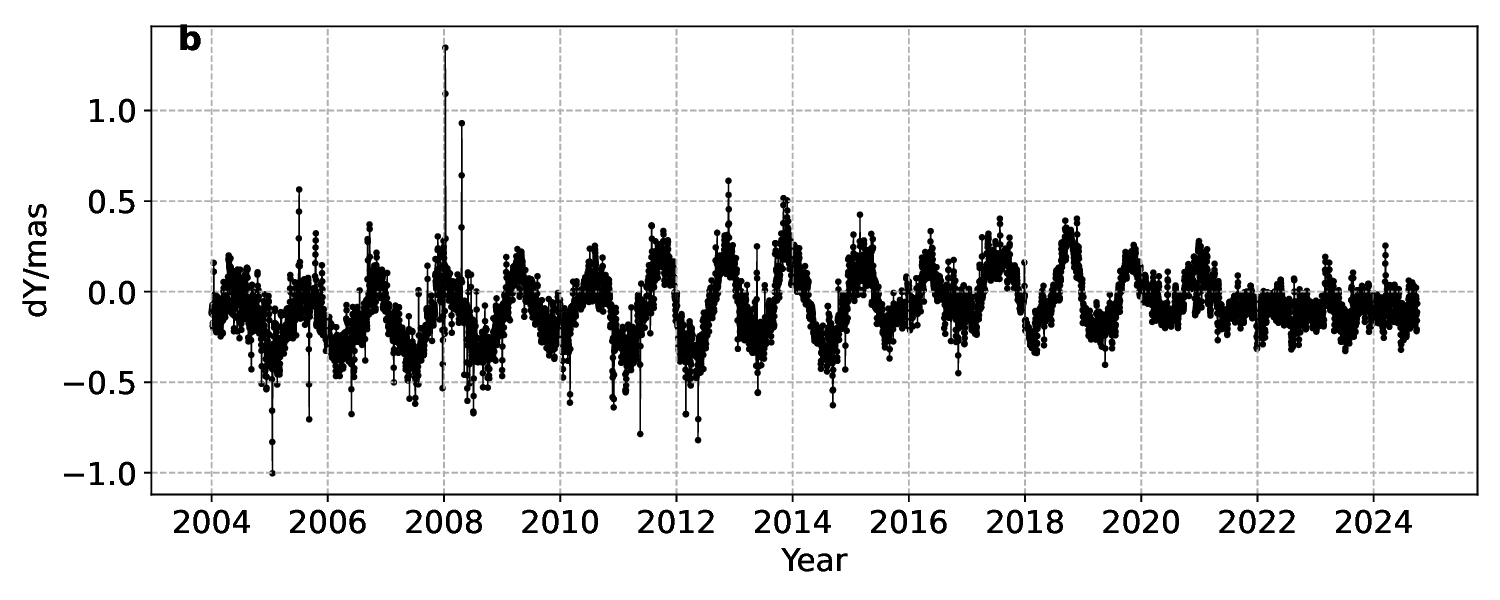}
			\caption{dY}
		\end{subfigure}
		\caption{Initial time series display of CPO 2002--2022 EOP 14 C04.}
		\label{fig:fig1}
	\end{figure}

	Since February 14, 2023, the International Terrestrial Reference Frame (ITRF) has been updated from ITRF 2014 to ITRF 2020. To maintain consistency with the Earth's reference frame, the EOP C04 series has also been updated from the 14th series to the 20th series. To evaluate the effectiveness of the CPO forecasting method presented in this paper, we adopted the EOP 14 C04 series for comparison with the CPO forecasting method in the 2nd EOP PCC. Additionally, to test the effectiveness of this method for the 20th series, we also demonstrated the applicability of this method under the EOP 14 C04 and EOP 20 C04 series in the discussion section. The initial time series of CPO is shown in Figure~\ref{fig:fig1}.
	
	\subsection{Bivariate least squares algorithm}
	
	The method employed in this paper is a parameter fitting approach based on sliding window bivariate least squares, enabling simultaneous modeling of the dX and dY signals of CPO. Song Yongzhi et al. utilized a similar method to achieve joint solution research on polar motion prediction methods, effectively enhancing the prediction accuracy of polar motion~\cite{SongYuChen2016PolarMotionJointPrediction}. Initially, the selection of data to be fitted is determined by the value of the fixed window, followed by estimating the parameter values through bivariate least squares fitting. By inputting the prediction length, the predicted portion is obtained. The bivariate least squares fitting formulas are presented in Equations~\eqref{eq:dx_model} and \eqref{eq:dy_model}.
	
	\begin{equation}
		\label{eq:dx_model}
		\mathrm{d}X = a + bt + \sum_{i=1}^{2}\left[As_i\sin\left(\frac{2\pi t}{T_i}\right)-Ac_i\cos\left(\frac{2\pi t}{T_i}\right)\right]+\mathrm{res}(\mathrm{d}X)
	\end{equation}
	
	\begin{equation}
		\label{eq:dy_model}
		\mathrm{d}Y = c + dt + \sum_{i=1}^{2}\left[Ac_i\sin\left(\frac{2\pi t}{T_i}\right)+As_i\cos\left(\frac{2\pi t}{T_i}\right)\right]+\mathrm{res}(\mathrm{d}Y)
	\end{equation}
	
	There $t$ is the mjd value corresponding to the CPO time series, $a$ and $c$ correspond to the integer parts of the dX and dY decompositions, respectively, $b$ and $d$ correspond to the trend term coefficients of the dX and dY decompositions, $As_i$ and $Ac_i$ are the amplitudes of the periodic terms, and $T_1$ and $T_2$ are the periods of the annual term (365 d) and the FCN signal (430 d), respectively. When selecting N CPO values, the joint observation equation can be written in matrix form as shown in Equation~\eqref{eq:joint_observation}:
	
	\begin{equation}
		\label{eq:joint_observation}
		L=Bp+v.
	\end{equation}
	
	There,
	\begin{equation}
		\label{eq:matrices}
		L=\begin{bmatrix}
			X(t_1)\\
			\vdots\\
			X(t_N)\\
			Y(t_1)\\
			\vdots\\
			Y(t_N)
		\end{bmatrix}_{2N\times1},\quad
		p=\begin{bmatrix}
			a\\
			b\\
			c\\
			d\\
			As_1\\
			Ac_1\\
			As_2\\
			Ac_2
		\end{bmatrix}_{8\times1},
	\end{equation}
	\begin{equation}
		\label{eq:B_matrix}
		B=\begin{bmatrix}
			1 & t_1 & 0 & 0 & c_{11} & -s_{11} & c_{12} & -s_{12}\\
			\vdots & \vdots & \vdots & \vdots & \vdots & \vdots & \vdots & \vdots\\
			1 & t_N & 0 & 0 & c_{N1} & -s_{N1} & c_{N2} & -s_{N2}\\
			0 & 0 & 1 & t_1 & s_{11} & c_{11} & s_{12} & c_{12}\\
			\vdots & \vdots & \vdots & \vdots & \vdots & \vdots & \vdots & \vdots\\
			0 & 0 & 1 & t_N & s_{N1} & c_{N1} & s_{N2} & c_{N2}
		\end{bmatrix}_{2N\times8}.
	\end{equation}
	$p$ is the parameter matrix of the joint model; $B$ is the coefficient matrix when solving the joint model; $L$ is the matrix consisting of observed values of CPO sequences dX and dY.
	
	The solution can be obtained as shown in Equation~\eqref{eq:least_squares_solution} based on the component residuals and the principle of minimization, $V^{T}V=\min$:
	\begin{equation}
		\label{eq:least_squares_solution}
		\hat{p}=\left(B^{T}B\right)^{-1}B^{T}L.
	\end{equation}
	The parameter values obtained from Equation~\eqref{eq:least_squares_solution} can be used to derive the joint solution model for determining the terms of the celestial pole deviation sequence.
	
	\subsection{Determination of sliding window length}
	
	Due to the time-varying intensity of FCN signals, the selection of the length of the sliding window is crucial for forecasting: Lambert's model uses a two-year fitting window, fits based on two years of data, and updates once a year~\cite{LambertDehant2007CoreParametersVLBI}; Belda confirmed the optimal sliding window of 400 days through extensive experiments~\cite{Belda2016FCNEmpiricalModel}; the ZM3 model proposed by Malkin uses an operational interval with a nominal FCN period of about 431 days, moving once a day~\cite{Malkin2013FCNGeomagneticJerks}.
	
	\begin{figure}[htbp]
		\centering
		\safeincludegraphics[width=0.85\linewidth]{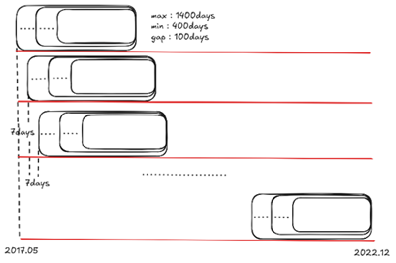}
		\caption{Schematic diagram of sliding window design.}
		\label{fig:fig2}
	\end{figure}
	
\begin{figure}[htbp]
		\centering
		\begin{subfigure}[t]{0.32\linewidth}
			\centering
			\safeincludegraphics[width=\linewidth]{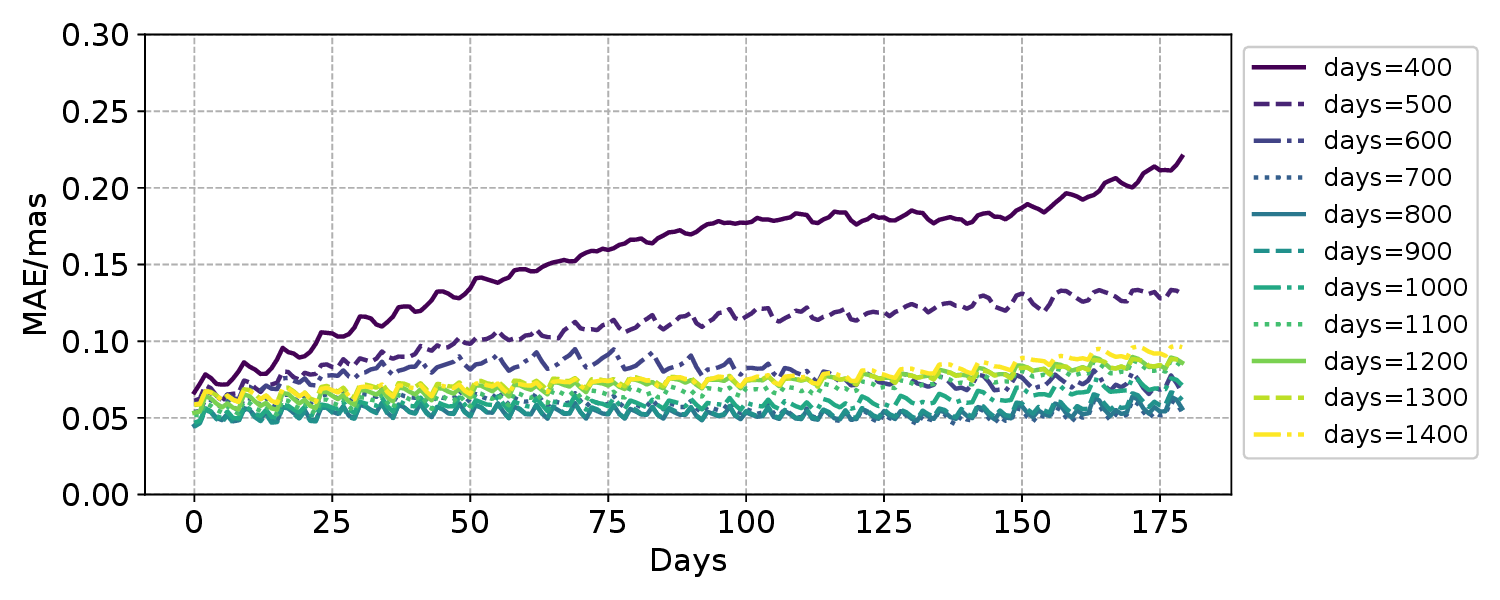}
			\caption{dX}
		\end{subfigure}\hfill
		\begin{subfigure}[t]{0.32\linewidth}
			\centering
			\safeincludegraphics[width=\linewidth]{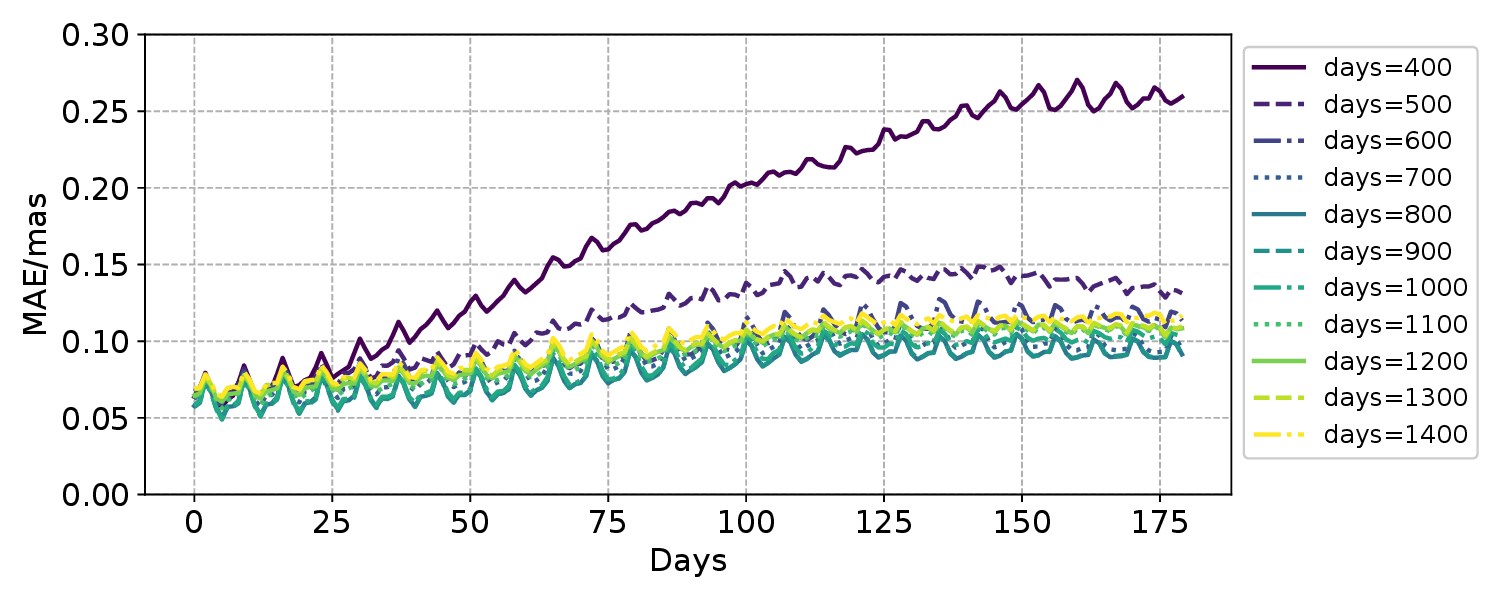}
			\caption{dY}
		\end{subfigure}\hfill
		\begin{subfigure}[t]{0.32\linewidth}
			\centering
			\safeincludegraphics[width=\linewidth]{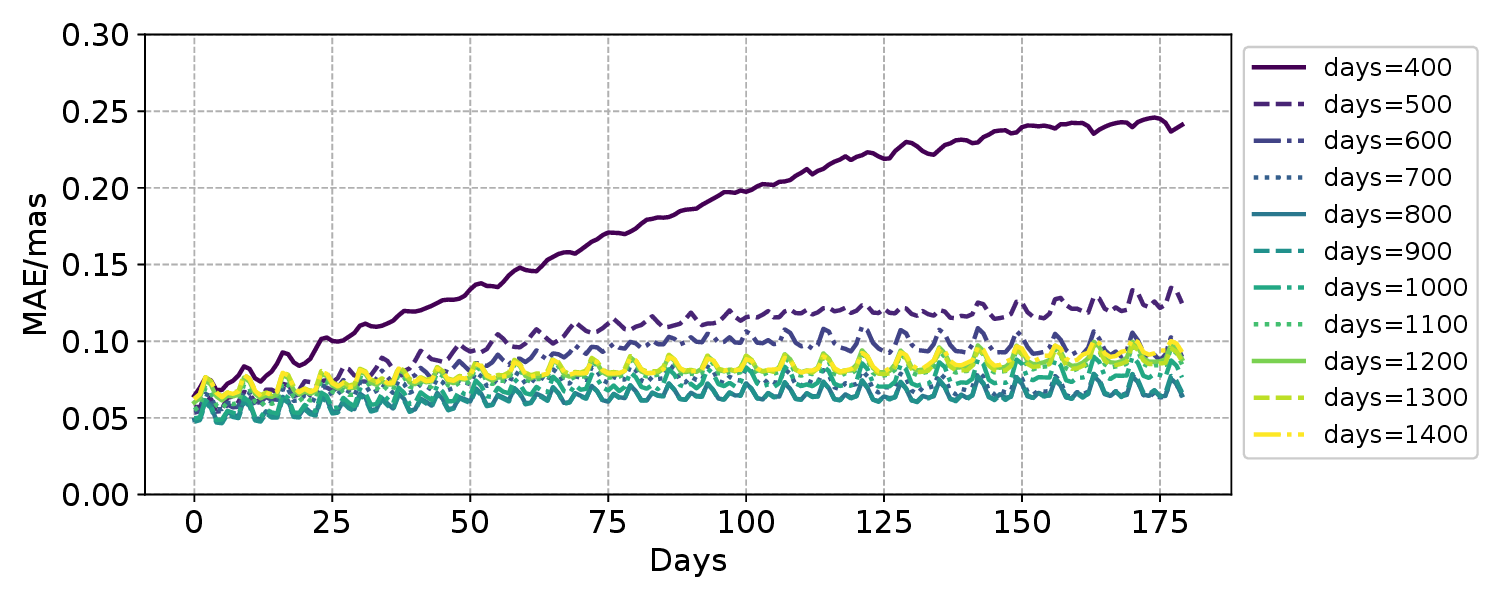}
			\caption{$\sqrt{\mathrm{d}X^{2}+\mathrm{d}Y^{2}}$}
		\end{subfigure}
		\caption{Comparison results of sliding windows.}
		\label{fig:fig3}
	\end{figure}

	Based on the comprehensive review of the aforementioned studies and the experience in fitting periodic signals, given that the main components have periods of 365 days and 430 days, it is generally believed that a length exceeding two periods should be selected. To determine the optimal sliding window, we conducted an algorithm experiment to select the best sliding window. As shown in Figure~\ref{fig:fig2}, data spanning 5.5 years from July 2017 to December 2022 was used. Forecasts were made every 7 days, with a total of 71 forecasts conducted. The prediction window moved once a day, and the sliding window interval was 100 days. Relevant experiments were conducted sequentially from 400 to 1400 days, comparing the MAE of forecasts under different window lengths. The experimental results are shown in Figure~\ref{fig:fig3}.
	
	As can be seen from Figure~\ref{fig:fig3}, whether considering single variables dX and dY or both variables comprehensively, the MAE performs well and is difficult to distinguish when the sliding window is between 700 and 1100.
	
	To further determine the optimal sliding window value, it is necessary to consider the performance in both dX and dY directions. Therefore, Table~\ref{tab:window_mae} provides detailed information on the MAE of $\sqrt{\mathrm{d}X^{2}+\mathrm{d}Y^{2}}$ for the 10th, 30th, 90th, and 180th days corresponding to 700-1100 days:
	
	\begin{table}[htbp]
		\centering
		\caption{MAE of $\sqrt{\mathrm{d}X^{2}+\mathrm{d}Y^{2}}$ under sliding window lengths of 700--1100.}
		\label{tab:window_mae}
		\begin{tabular}{lccccc}
			\toprule
			\multirow{2}{*}{Forecast span} & \multicolumn{5}{c}{MAE (mas)} \\
			\cmidrule(lr){2-6}
			& 700d & 800d & 900d & 1000d & 1100d \\
			\midrule
			10th day  & 0.066 & 0.059 & 0.058 & 0.060 & 0.069 \\
			30th day  & 0.073 & 0.066 & 0.065 & 0.069 & 0.076 \\
			90th day  & 0.077 & 0.066 & 0.066 & 0.072 & 0.079 \\
			180th day & 0.064 & 0.065 & 0.067 & 0.077 & 0.085 \\
			\bottomrule
		\end{tabular}
	\end{table}
	
	As can be seen from Table~\ref{tab:window_mae}, the MAE corresponding to 10, 30, and 90 days is the smallest, with only a difference of 3 uas compared to 600 days in the 180-day prediction. Taking all factors into consideration, we have chosen a fixed value of 900 days for the sliding window in this study.
	
\begin{figure}[htbp]
		\centering
		\safeincludegraphics[width=0.85\linewidth]{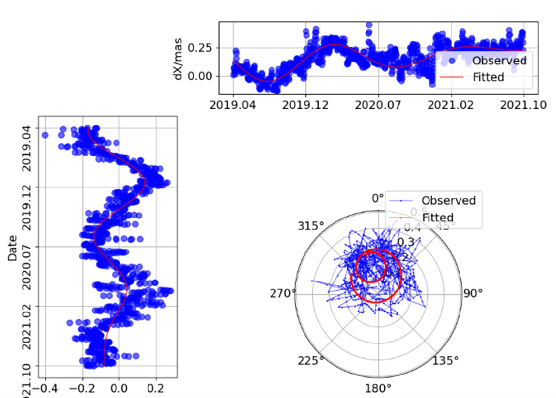}
		\caption{Single fitting result.}
		\label{fig:fig4}
	\end{figure}

	After selecting a sliding window of 900 days, this paper fits the selected 900-day data based on the bivariate least squares fitting algorithm presented in formula (1). Figure~\ref{fig:fig4} shows the single fitting result of the algorithm. From Figure~\ref{fig:fig4}, it can be seen that under the constraints of bivariate least squares, the fitting results for variables dX and dY are both good.
	
	\section{Evaluation results and discussion}
	
	To evaluate the accuracy of predicting dX and dY using the algorithm proposed in this paper, the widely used mean absolute error (MAE) is selected as the accuracy evaluation metric. MAE can be expressed as:
	
	\begin{equation}
		\label{eq:error}
		E_j=P_j^i-O_j^i.
	\end{equation}
	
	\begin{equation}
		\label{eq:mae}
		MAE_j=\frac{1}{n}\sum_{i=1}^{n}\left|P_j^i-O_j^i\right|,\quad i=1,\cdots,n.
	\end{equation}
	
	where $O$ is the predicted value of dX and dY, $P$ is the observed value of dX and dY, $i$ is mjd, $j$ is the predicted day number from 1 to 180, and $n$ represents the total number of predictions.
	
	\subsection{Presentation of forecast results using the algorithm in this paper}
	
\begin{figure}[htbp]
		\centering
		\begin{subfigure}[t]{0.48\linewidth}
			\centering
			\safeincludegraphics[width=\linewidth]{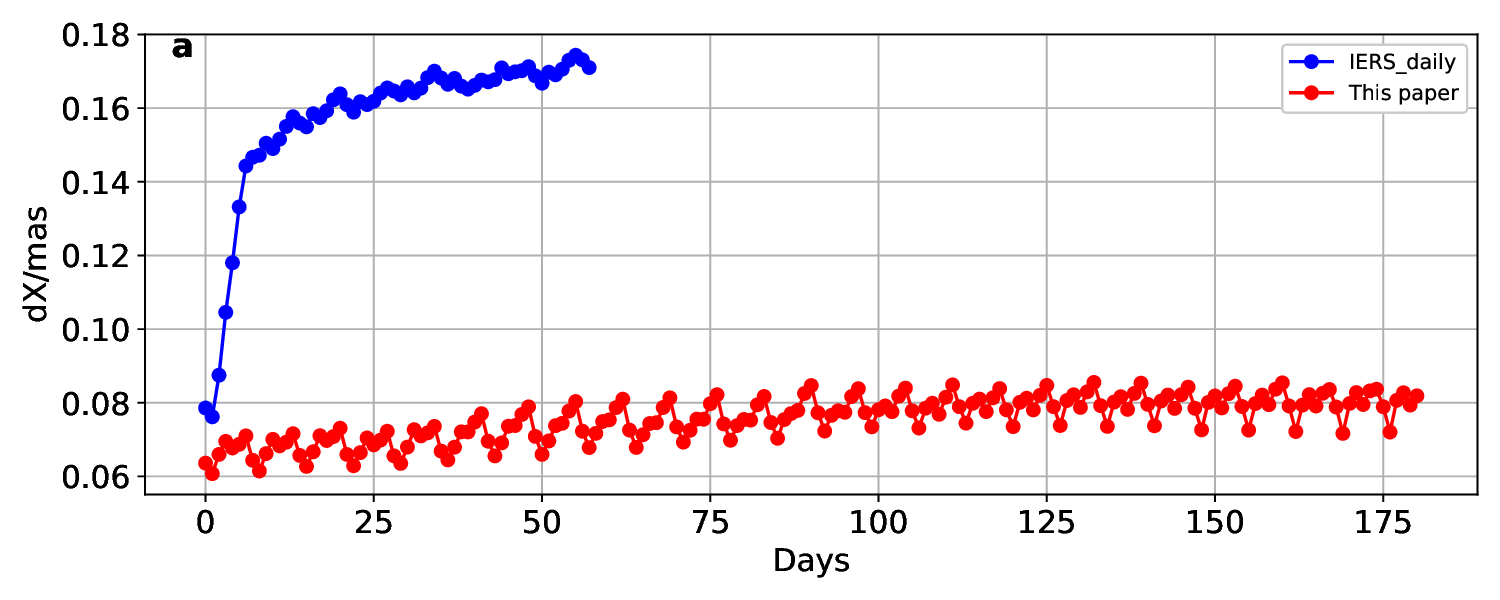}
			\caption{dX}
		\end{subfigure}\hfill
		\begin{subfigure}[t]{0.48\linewidth}
			\centering
			\safeincludegraphics[width=\linewidth]{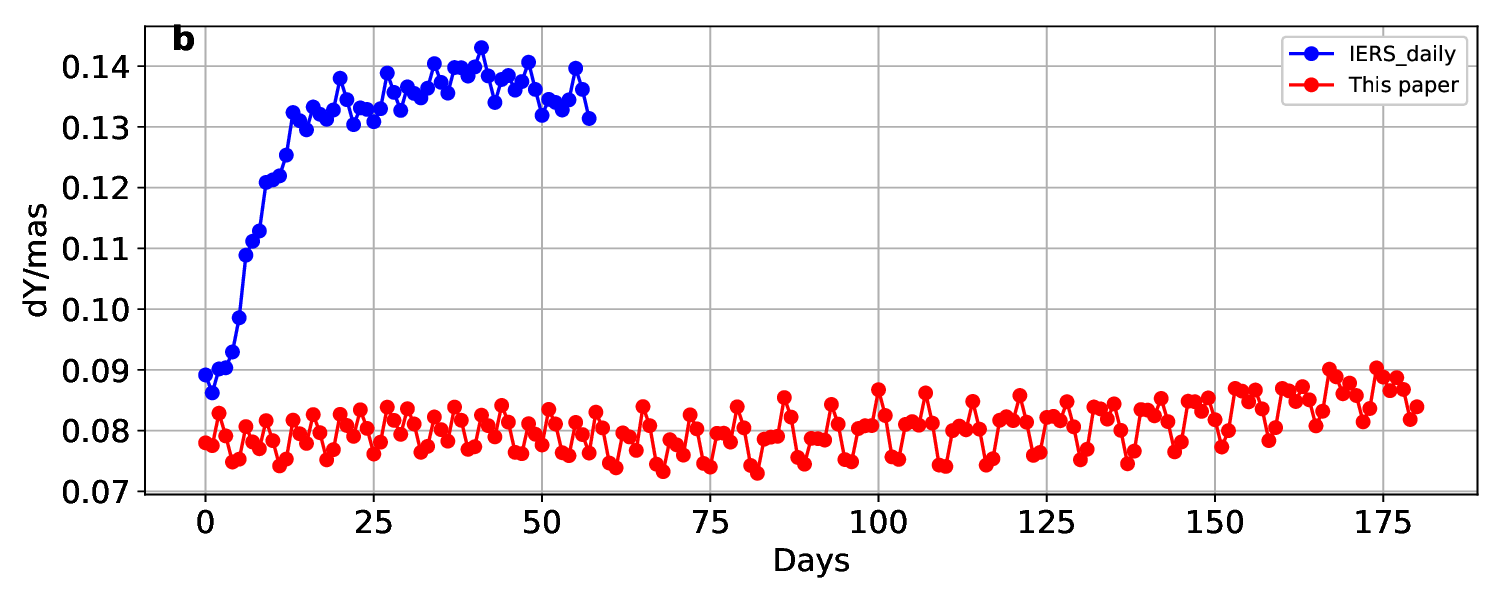}
			\caption{dY}
		\end{subfigure}
		\caption{Comparison of forecast results.}
		\label{fig:fig5}
	\end{figure}

	To verify the generalization ability of the algorithm proposed in this paper across different time spans, based on the sliding window derived from 5.5 years of data from July 2017 to December 2022, the following experiment demonstrates the forecasting performance of the algorithm in a two-year time span from March 2023 to March 2025, with a total of 109 forecasting experiments conducted once a week. To enhance the persuasiveness of the experimental results, the daily files published by IERS were again used as a control experiment. The daily files provided by IERS offer rapid solutions and forecasts for parameters such as polar motion, UT1--UTC, and celestial pole deviation on a daily basis. The CPO provides a 90-day forecast starting from the date of the rapid solution completion, which means that the actual number of days forecasted backwards is less than 90 days. The results of this control experiment are shown in Figure~\ref{fig:fig5}.
	
	Based on the comparative experiment in Figure~\ref{fig:fig5}, it can be seen that under a prediction span of 180 days, the accuracy of dX predicted by the algorithm in this paper is 61-73 uas for 0-30 days, 64-80 uas for 30-57 days, and 68-86 uas for 57-180 days, which is significantly better than the performance of the daily files from IERS during the same period, which is 76-166 uas for 0-30 days and 164-174 uas for 30-57 days. The average prediction errors for the 10th, 30th, and 57th days decreased by 53\%, 59\%, and 60\%, respectively. The accuracy of dY predicted by the algorithm in this paper is 74-84 uas for 0-30 days, 76-84 uas for 30-57 days, and 73-90 uas for 57-180 days, which is significantly better than the performance of the daily files from IERS during the same period, which is 86-139 uas for 0-30 days and 131-143 uas for 30-57 days. The average prediction errors for the 10th, 30th, and 57th days decreased by 35\%, 38\%, and 42\%, respectively.
	
\begin{figure}[htbp]
		\centering
		\begin{subfigure}[t]{0.48\linewidth}
			\centering
			\safeincludegraphics[width=\linewidth]{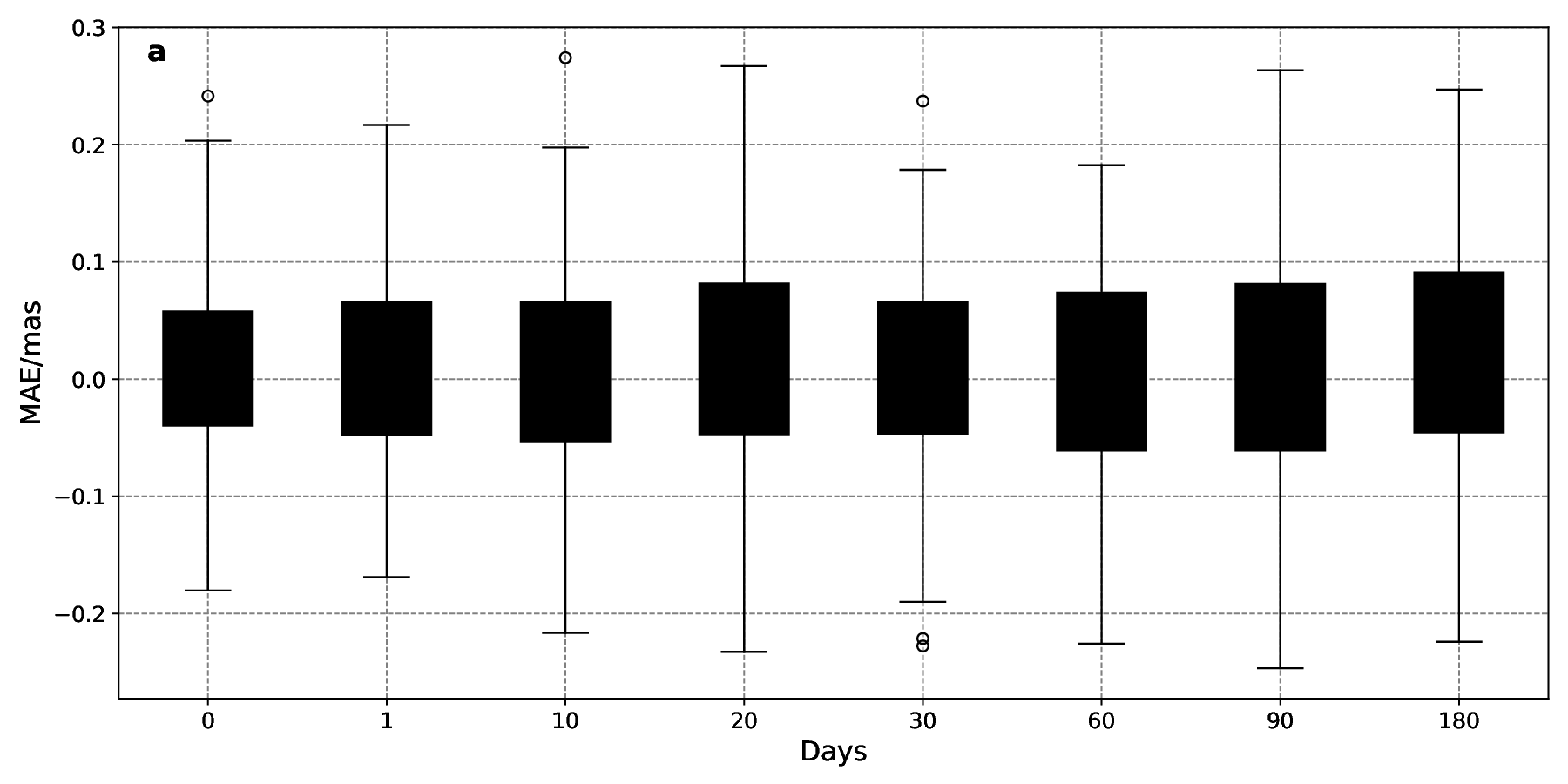}
			\caption{dX}
		\end{subfigure}\hfill
		\begin{subfigure}[t]{0.48\linewidth}
			\centering
			\safeincludegraphics[width=\linewidth]{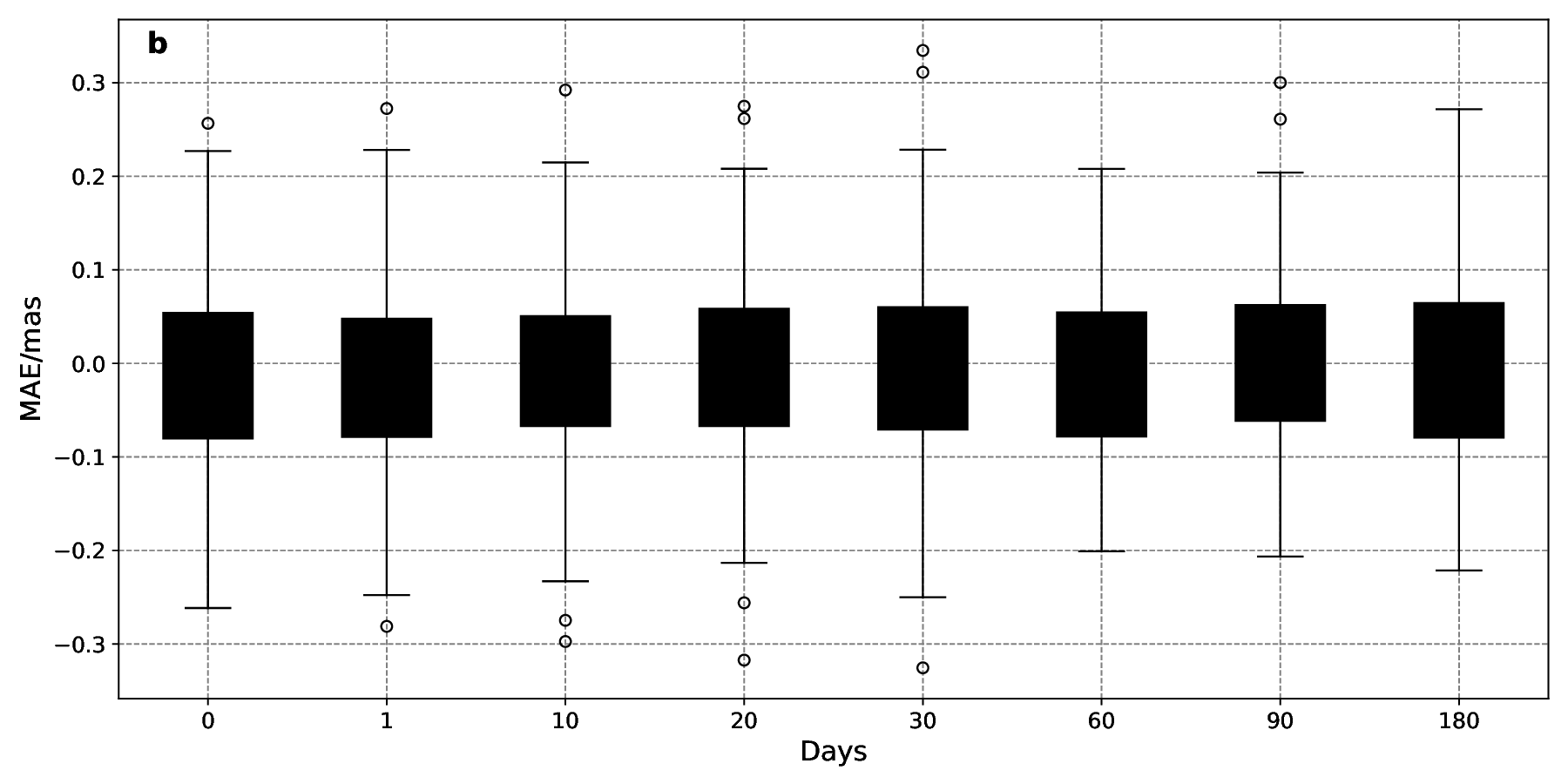}
			\caption{dY}
		\end{subfigure}
		\caption{Box plot of forecast error.}
		\label{fig:fig6}
	\end{figure}

	To reflect the credibility of this experiment, Figure~\ref{fig:fig6} presents a box plot of the forecast errors, which are the differences between the forecast results and the actual values for 0-180 days across 109 forecasts. As can be seen from the figure, the errors of dX and dY are generally concentrated around 0, indicating that the deviations between the forecast results and the actual values are small, and the overall forecast accuracy is high. In terms of dispersion, the box shapes (interquartile ranges) of both are relatively narrow, indicating that the error distribution across the 109 experiments is relatively concentrated, with no large-scale fluctuations; and the small number of outliers beyond the upper and lower whiskers in the figure accounts for a relatively low proportion, further reflecting the stability and reliability of the experimental results. Combining these two characteristics of the box plots of forecast errors, it can be concluded that the 109 forecast results of this experiment exhibit good consistency and credibility.
	
	\subsection{Comparison of CPO forecast results between 2.2 and 2nd EOP PCC}
	
	The 2nd EOP PCC has only been held for over a year, which is insufficient to evaluate long-time series forecasts. However, due to CPO is a variable with relatively slow changes, the evaluation of long-time series forecasts has certain practical significance. Currently, with the accumulation of CPO time series, there are conditions for evaluating long-time forecasts. Therefore, this paper incorporates the results of long-time series forecasts to further verify the forecasting ability of the algorithm presented in this paper.
	
	To verify the method proposed in this paper, a total of 70 prediction experiments were conducted at a frequency of once per week within the time span from September 1, 2021, to December 31, 2022. These predictions were compared with those made by the team with the best performance in the second EOP PCC, as well as the files uploaded by our team during the same period and the daily files of IERS; our team's ID is 101, and the method used is univariate LS+AR; the group with the best dX prediction has an ID of 154, and the group with the best dY prediction has an ID of 155, both using first-order neural ordinary differential equations.
	
\begin{figure}[htbp]
		\centering
		\begin{subfigure}[t]{0.48\linewidth}
			\centering
			\safeincludegraphics[width=\linewidth]{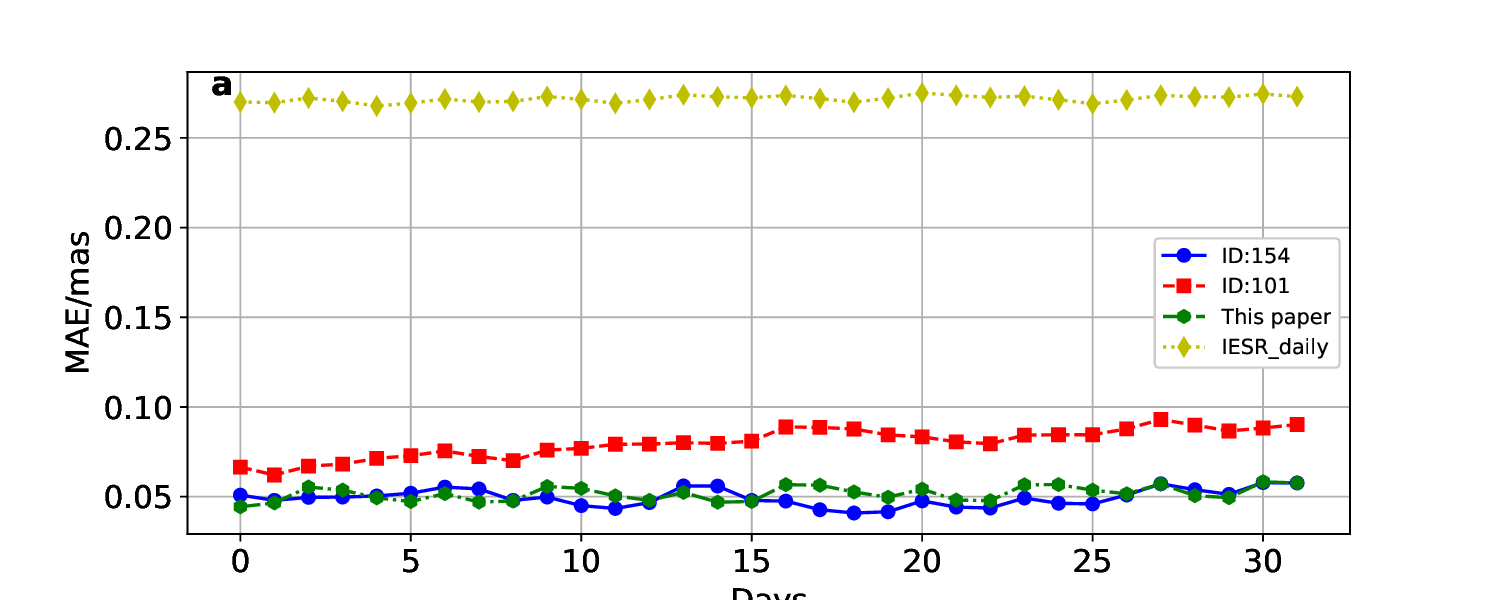}
			\caption{30 days}
		\end{subfigure}\hfill
		\begin{subfigure}[t]{0.48\linewidth}
			\centering
			\safeincludegraphics[width=\linewidth]{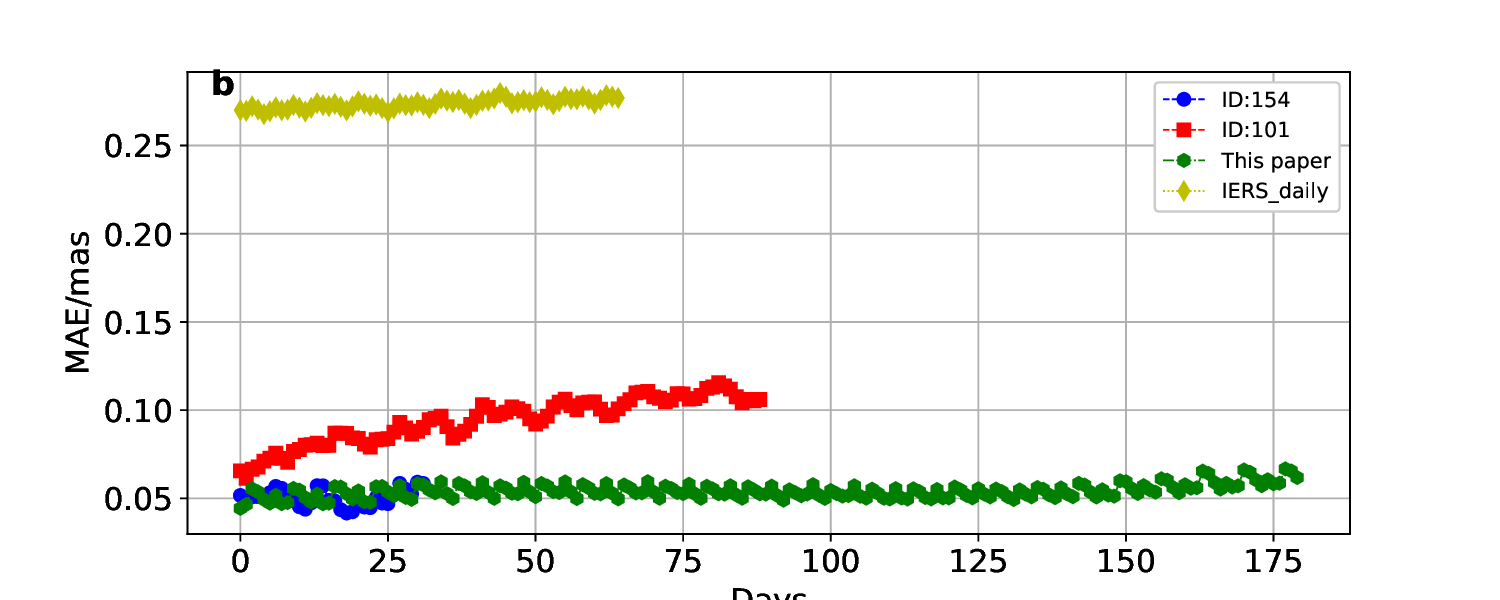}
			\caption{180 days}
		\end{subfigure}
		\caption{Comparison of dX forecast results.}
		\label{fig:fig7}
	\end{figure}

\begin{figure}[htbp]
		\centering
		\begin{subfigure}[t]{0.48\linewidth}
			\centering
			\safeincludegraphics[width=\linewidth]{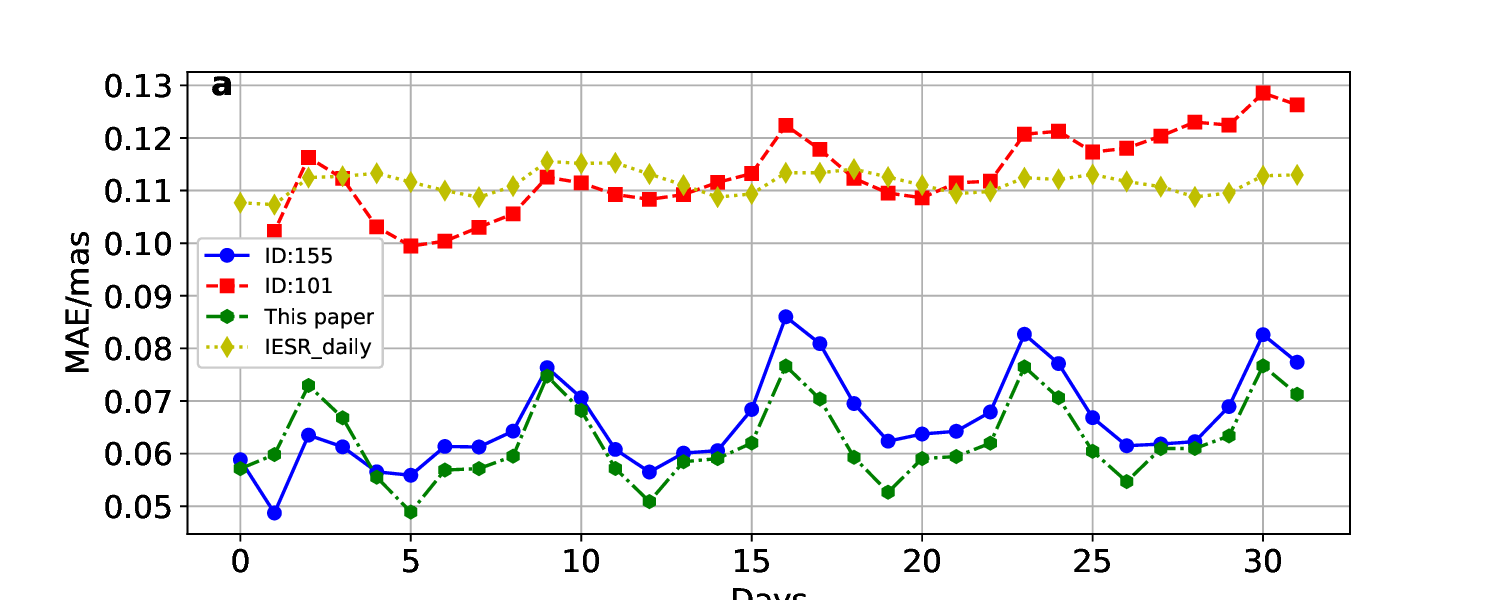}
			\caption{30 days}
		\end{subfigure}\hfill
		\begin{subfigure}[t]{0.48\linewidth}
			\centering
			\safeincludegraphics[width=\linewidth]{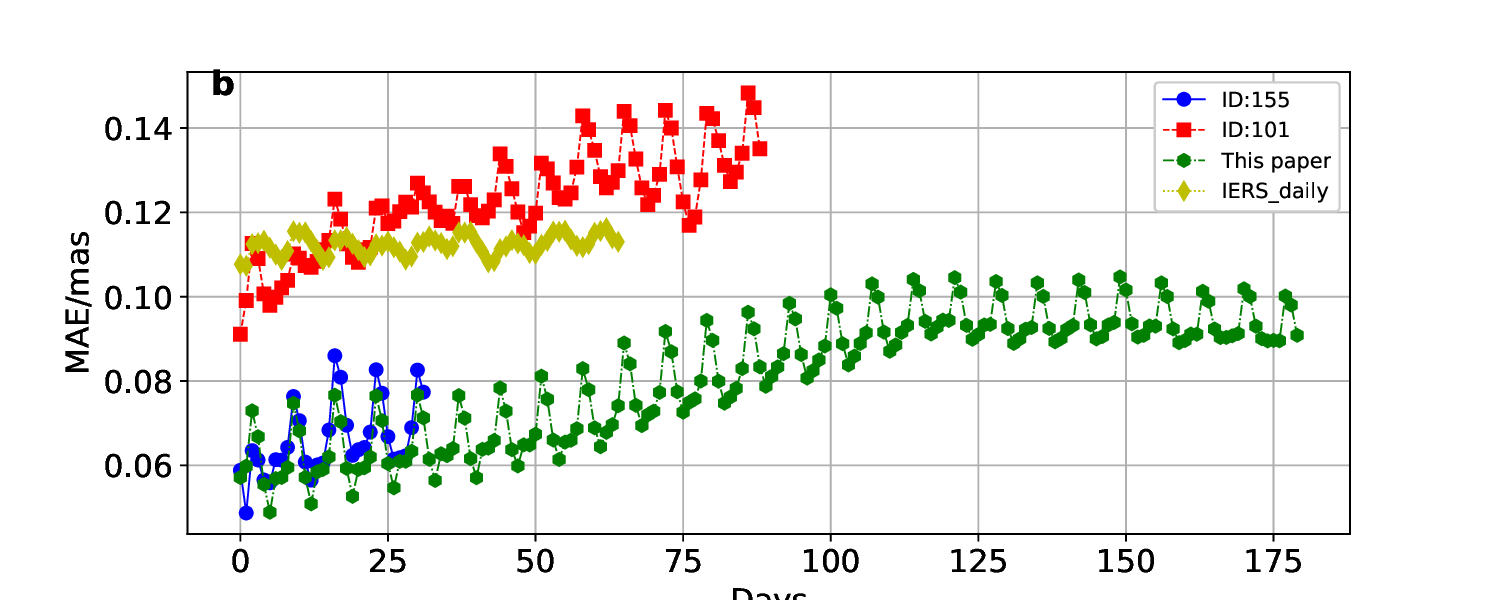}
			\caption{180 days}
		\end{subfigure}
		\caption{Comparison of dY forecast results.}
		\label{fig:fig8}
	\end{figure}

	The MAE obtained by comparing these files with the EOP C04 files published by IERS is shown in Figures~\ref{fig:fig7} and \ref{fig:fig8}.
	
	As can be seen from the figure, the method we used this time has achieved overall improvement compared to the method used in the 2nd EOP PCC, and is superior to the daily files from IERS. For the team's forecast results during the 2nd EOP PCC, in the dX direction, the average forecast errors for days 7, 30, and 88 decreased by 35\%, 34\%, and 51\%, respectively; in the dY direction, the average forecast errors for days 7, 30, and 88 decreased by 45\%, 40\%, and 38\%, respectively. For the forecast results from the daily files, in the dX direction, the average forecast errors for days 7, 30, and 60 decreased by 83\%, 80\%, and 81\%, respectively; in the dY direction, the average forecast errors for days 7, 30, and 60 decreased by 48\%, 32\%, and 40\%, respectively. The forecast for dX is comparable to the best performance of the 2nd EOP PCC, and is not inferior to it on days 0, 1, 7, and 31; the forecast for dY is only inferior to the best performance of the 2nd EOP PCC from day 1 to day 3, and is superior to it on days 0 and from day 4 to day 31. The specific MAE numerical statistical results for the dX and dY directions are shown in Table~\ref{tab:dx_results} and Table~\ref{tab:dy_results}, respectively:
	
	\begin{table}[htbp]
		\centering
		\caption{Comparison of dX forecast results.}
		\label{tab:dx_results}
		\begin{tabular}{lcccc}
			\toprule
			\multirow{2}{*}{Forecast spans} & \multicolumn{4}{c}{MAE (mas)} \\
			\cmidrule(lr){2-5}
			& ID:154 & ID:101 & This paper & IERS\_daily \\
			\midrule
			0d   & 0.051 & 0.066 & 0.044 & 0.270 \\
			1d   & 0.048 & 0.062 & 0.047 & 0.270 \\
			7d   & 0.054 & 0.072 & 0.047 & 0.270 \\
			15d  & 0.048 & 0.081 & 0.047 & 0.272 \\
			30d  & 0.058 & 0.088 & 0.058 & 0.275 \\
			31d  & 0.058 & 0.090 & 0.058 & 0.273 \\
			60d  & /     & 0.105 & 0.053 & 0.274 \\
			88d  & /     & 0.106 & 0.052 & /     \\
			180d & /     & /     & 0.062 & /     \\
			\bottomrule
		\end{tabular}
	\end{table}
	
	\begin{table}[htbp]
		\centering
		\caption{Comparison of dY forecast results.}
		\label{tab:dy_results}
		\begin{tabular}{lcccc}
			\toprule
			\multirow{2}{*}{Forecast spans} & \multicolumn{4}{c}{MAE (mas)} \\
			\cmidrule(lr){2-5}
			& ID:155 & ID:101 & This paper & IERS\_daily \\
			\midrule
			0d   & 0.059 & 0.094 & 0.057 & 0.108 \\
			1d   & 0.049 & 0.102 & 0.060 & 0.107 \\
			7d   & 0.061 & 0.103 & 0.057 & 0.109 \\
			15d  & 0.068 & 0.113 & 0.062 & 0.109 \\
			30d  & 0.083 & 0.129 & 0.077 & 0.113 \\
			31d  & 0.077 & 0.126 & 0.071 & 0.113 \\
			60d  & /     & 0.135 & 0.069 & 0.115 \\
			88d  & /     & 0.135 & 0.084 & /     \\
			180d & /     & /     & 0.091 & /     \\
			\bottomrule
		\end{tabular}
	\end{table}
	
	From Table~\ref{tab:dx_results}, it can be observed that the dX forecast results are optimal for the current algorithm on Day 0, Day 1, Day 7, and Day 15. On Day 30 and Day 31, the best results of the current algorithm are identical to those of the 2nd EOP PCC. Table~\ref{tab:dy_results} reveals that on Day 1, the dY forecast results of the optimal group of the 2nd EOP PCC are optimal. However, on Day 0, Day 7, Day 15, Day 30, and Day 31, the forecast results of the current algorithm are optimal. Additionally, the forecast results of the current algorithm after Day 31 are relatively stable.
	
	\subsection{Comparison between EOP 14 C04 sequence and EOP 20 C04 sequence}
	
	The results of the 2nd EOP PCC competition indicate that the EOP 20 C04 sequence has little impact on UT1 and polar motion compared to the EOP 14 C04 sequence, but it has a significant impact on CPO~\cite{Winska2024SecondEOPPCC_CPO}. To verify this competition conclusion and test the generalization ability of the algorithm in this paper on different sequences, this paper added comparative experiments between the algorithm and the EOP 14 C04 sequence and EOP 20 C04 sequence.
	
	The 2nd EOP PCC was held from September 2021 to December 2022. During this period, the EOP 14 C04 sequence was not converted to the EOP 20 C04 sequence. Therefore, only the daily files of the EOP 14 C04 sequence downloaded from the official website were stored, and there were no daily files of the EOP 20 C04 sequence. Based on this situation, when verifying the impact of the algorithm in this paper on the two sequences, we actually compared the concatenated daily files of the EOP 14 C04 sequence with the EOP 20 C04 sequence of corresponding dates.
	
	\begin{figure}[htbp]
		\centering
		\begin{subfigure}[t]{0.48\linewidth}
			\centering
			\safeincludegraphics[width=\linewidth]{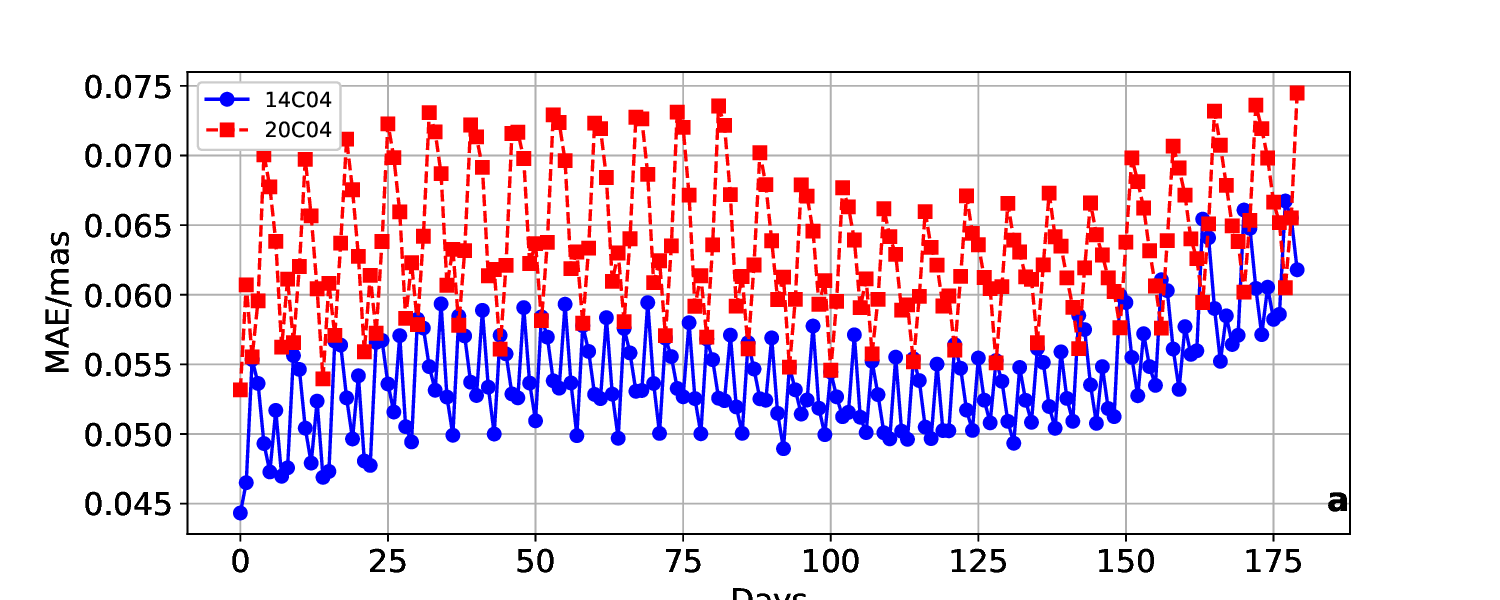}
			\caption{dX}
		\end{subfigure}\hfill
		\begin{subfigure}[t]{0.48\linewidth}
			\centering
			\safeincludegraphics[width=\linewidth]{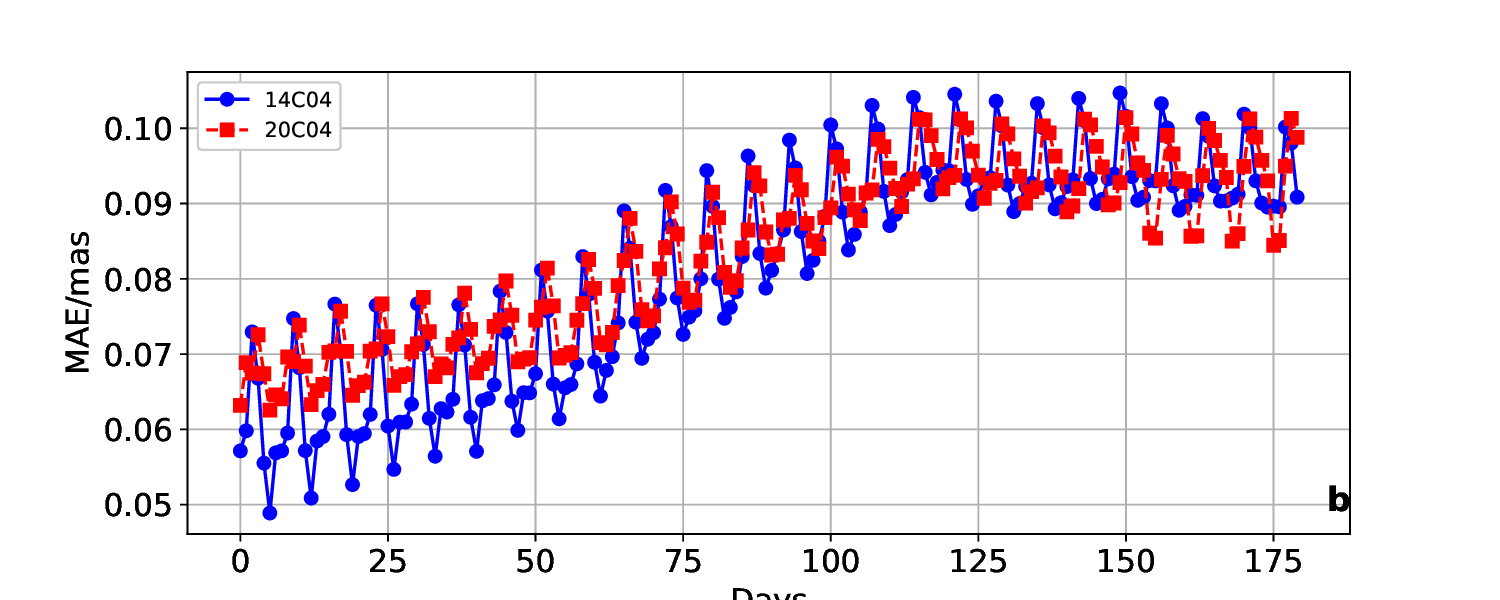}
			\caption{dY}
		\end{subfigure}
		\caption{Comparison between EOP 14 C04 and EOP 20 C04.}
		\label{fig:fig9}
	\end{figure}
	
	As can be seen from Figure~\ref{fig:fig9}, different sequences indeed have an impact on the forecasting of CPO. In the dX direction, the sequence effect is more pronounced. Compared to the 20C04 sequence, the 14C04 sequence reduces the average forecasting errors by 12\%, 0\%, 11\%, and 17\% for the 10th, 30th, 90th, and 180th days, respectively. In the dY direction, the short- and medium-term sequence effects are more significant. The 20C04 sequence is generally weaker than the 14C04 sequence for the first 60 days. Compared to the 20C04 sequence, the 14C04 sequence reduces the average forecasting errors by 8\%, -7\%, and 12\% for the 10th, 30th, and 60th days, respectively. However, in the medium- and long-term range of 61-180 days, the two sequences are basically equivalent, and the forecasting differences between them cannot be clearly distinguished.
	
	The CPO forecast results for the EOP 14 C04 series are superior to those for the EOP 20 C04 series, indicating that the consistency between the EOP 20 C04 series and ITRF2020 is not as good as that between the EOP 14 C04 series and ITRF2014. The reason behind this may stem from the imperfections of the Precession-Nutation model (PN model) for the 2000/2006A epoch. Regarding the differences in CPO prediction results between the EOP 14 C04 and EOP 20 C04 series, the difference in the dX component is more pronounced, which is also consistent with the data presented in Figure~\ref{fig:fig1}. It is evident from Figure~\ref{fig:fig1} that there is a non-stationary upward drift in the long-term trend of the dX series. This is due to the use of a fixed geodynamic flattening in the 2000/2006 PN model, whereas recent second-degree harmonic coefficient J2 data for the Earth's gravitational field indicate that J2 and geodynamic flattening are not constants but exhibit parabolic-like time variability~\cite{LiuHuang2025IAU2006J2}. Therefore, re-solving VLBI data using a new PN model and updating CPO data based on the new model are expected to enhance the consistency between the EOP series and the ITRF framework, as well as improve the forecast accuracy of CPO parameters.
	
	\subsection{Comparison of different methods for handling least squares residuals}
	
	We found that when the level of CPO least squares residuals is very low, the extrapolation prediction accuracy is better when only using least squares fitting without any processing of the residuals.
	
	\begin{figure}[htbp]
		\centering
		\begin{subfigure}[t]{0.48\linewidth}
			\centering
			\safeincludegraphics[width=\linewidth]{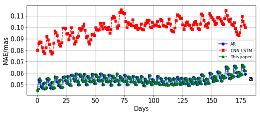}
			\caption{dX}
		\end{subfigure}\hfill
		\begin{subfigure}[t]{0.48\linewidth}
			\centering
			\safeincludegraphics[width=\linewidth]{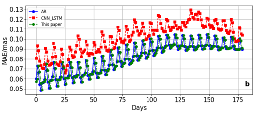}
			\caption{dY}
		\end{subfigure}
		\caption{MAE corresponding to different residual processing methods.}
		\label{fig:fig10}
	\end{figure}
	
	Figure~\ref{fig:fig10} compares the MAE of residuals without processing, residuals fitted by AR, and residuals fitted by Convolutional Neural Network-Long Short-Term Memory (CNN-LSTM). The results show that there is almost no difference between the MAE of residuals without processing and the MAE after AR processing.
	
	We provide the following explanation for this phenomenon: Through the Augmented Dickey-Fuller (ADF) test and Bayesian Information Criterion (BIC) order determination analysis of the bivariate least squares residuals, it was found that the residual sequence is close to a white noise distribution (BIC=3). Meanwhile, the actual prediction results show that the AR model can only provide a weak correction effect within a short time horizon (7-15 days). As the time horizon increases, the predicted values quickly converge to a constant straight line near 0. However, in this paper, the initial CPO sequence of C04 superimposed with daily rapid solution starts the actual forecast about 20 days before the application forecast start date. That is, the contribution of the residual AR model after the application forecast start date is essentially 0, which is consistent with Figure~\ref{fig:fig10}. Considering the computational complexity of the model and the stability of long-term forecasting, this paper ignores residual correction during the long-term extrapolation stage and sets it directly to 0 to avoid introducing additional parameter estimation risks.
	
	This phenomenon is only observed in the forecast of CPO and does not hold for UT1, PMX, or PMY. We believe the reason for this is that the CPO sequence updates slowly, and CPO changes gradually. When the least squares model already contains the main information, the high-frequency measurement noise in the fitting residuals becomes dominant.
	
	\section{Conclusion}
	
	Addressing the common issues in the CPO forecasting process during the 2nd EOP PCC, such as the difficulty in simultaneously considering two variables and the challenges in data fitting, this paper improves upon the existing algorithms of our team. We employ a sliding window combined with bivariate least squares algorithm to re-fit and predict the 70 datasets corresponding to the period from September 1, 2021, to December 31, 2022, during the 2nd EOP PCC. The results are then compared with the best performance during the competition, our team's performance during the competition, and the daily files. The findings indicate that the forecast results of dX using this algorithm are notably superior to those of our team and the daily files during this period, and also slightly better than the best forecast results of the 2nd EOP PCC. Furthermore, the forecast results of dY using this algorithm are overall superior to those of all groups in the 2nd EOP PCC and the daily files.
	
	Meanwhile, this paper experimentally demonstrates the conclusion from the 2nd EOP PCC competition that the 14 C04 and 20 C04 sequences have a significant impact on CPO. The experimental results show that, based on the application of the algorithm proposed in this paper, sequence changes have a notable effect on the dX component of CPO; however, for the dY component, sequence changes only have a slight impact in the early and middle stages from day 0 to day 60, and essentially have no impact in the medium and long-term stages from day 61 to day 180. This may be due to the imperfections of the PN model and the consistency issue between the EOP 20 C04 sequence and the ITRF2020 framework. In the future, research on CPO prediction based on a new PN model will be conducted to verify this hypothesis.
	
	Finally, to verify the generalization ability of the algorithm proposed in this paper across different time spans, a total of 109 experiments were conducted weekly over a two-year period from March 2023 to March 2025, following the conclusion of the 2nd EOP PCC. The experimental results demonstrate that the algorithm presented in this paper exhibits strong generalization capabilities, high accuracy in prediction results, and stability over a period of 0-180 days, significantly outperforming the CPO predictions from the daily files of the IERS during the same period.
	
	Regarding the fact that the bivariate least squares fitting prediction results used in this paper are significantly better than those predicted by the official daily files of IERS, this paper believes that it is due to the selection of an appropriate sliding window and the superior performance of the bivariate least squares algorithm in fitting the CPO sequence compared to other teams. At the same time, it is observed that the Lambert model adopted by IERS only considers FCN and disregards the linear trend of CPO, so the impact on ultra-short-term prediction is not significant, but the long-term prediction is less effective. These two factors combined lead to the above results.
	
	\section*{Thanks}
	
	Thank Professor Santiago Belda from The University of Alicante, Spain, for sharing the FCN model data and providing assistance with the CPO forecasting algorithm.
	
	This article utilizes the shared competition data from the Second International EOP Prediction Competition (\url{http://eoppcc.cbk.waw.pl}) and the EOP data from IERS (\url{https://www.iers.org/IERS/EN/DataProducts/EarthOrientationData/eop.html}).

\end{document}